# OS Debugging Method Using a Lightweight Virtual Machine Monitor

Tadashi Takeuchi
*Hitachi Systems Development Laboratory, Japan*
*t-takeu@sdl.hitachi.co.jp*

## Abstract

*Demands for implementing original OSs that can achieve high I/O performance on PC/AT compatible hardware have recently been increasing, but conventional OS debugging environments have not been able to simultaneously assure their stability, be easily customized to new OSs and new I/O devices, and assure efficient execution of I/O operations. We therefore developed a novel OS debugging method using a lightweight virtual machine. We evaluated this debugging method experimentally and confirmed that it can transfer data about 5.4 times as fast as the conventional virtual machine monitor.*

## 1. Introduction

Many appliance servers using original real-time operating systems (OSs) and achieving high-performance I/O on PC/AT compatible hardware have been marketed recently. One result of this trend is increasing demand for efficient debugging environments for original OSs to support various I/O devices and efficient debugging mechanisms monitoring the OS status tracing even while the OS is executing high-throughput I/O operations.

Because the vendors of PC/AT-compatible hardware do not provide debugging hardware such as ICE, the only debugging environments available for operating systems running on PCs are a hardware simulator working with a software debugger, a software remote debugger, and a software debugger embedded in the operating system under development. These environments, however, cannot simultaneously assure their stability when the operating system under development does not perform properly, be easily customized for different operating systems and new I/O devices, and assure efficient execution of I/O operations.

We therefore developed a novel OS debugging environment and method that can satisfy the above three demands simultaneously by using a lightweight virtual machine monitor. In this paper, we describe our new environment and method and report the results of experiments comparing its performance with that of conventional debugging environments.

## 2. Debugging Method

Our new debugging method provides the debugging environment shown in Fig. 2.1.

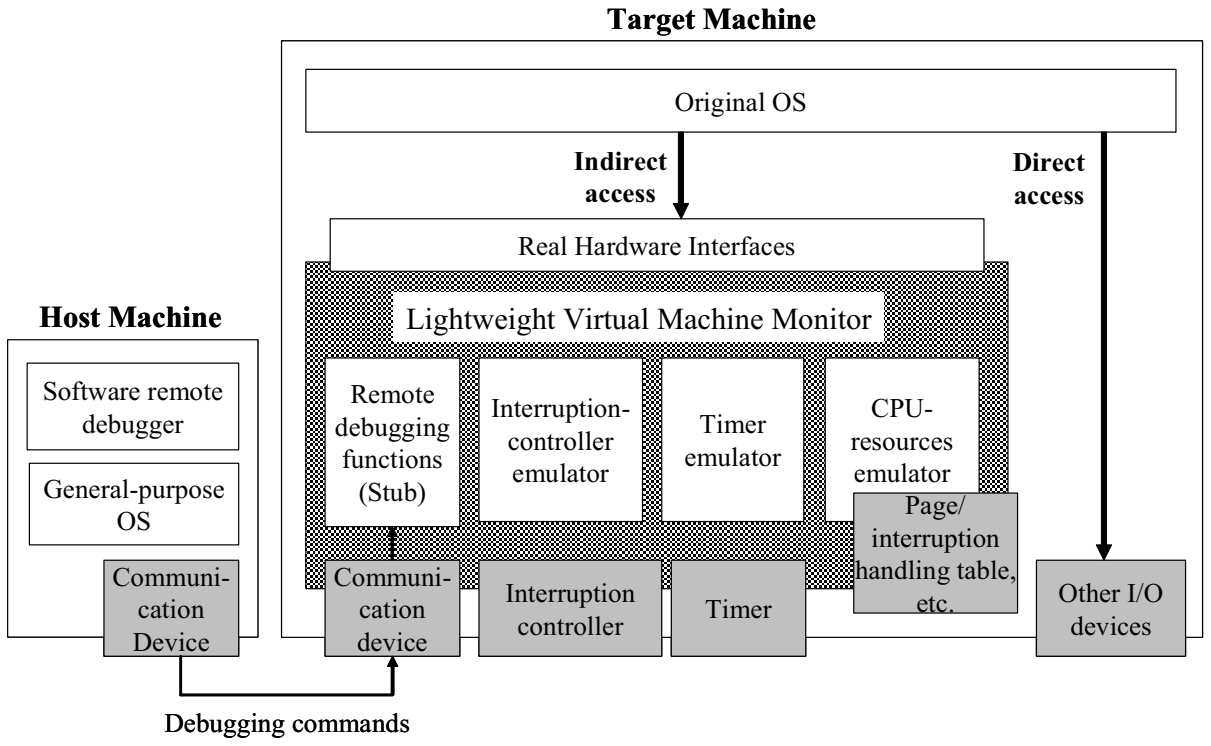

**Figure 2.1 Architecture of debugging environment**

The architecture of our debugging environment is similar to the one used for conventional remote debugging. It consists of a host machine and a target machine. A software remote debugger running on the host machine receives debugging commands (target machine memory/register reference/updating, etc.) from a user and sends the command to the target machine

The target machine architecture differs from that of classical software remote debugging in that a lightweight virtual machine monitor implemented independently of original OSs is embedded in it. The monitor provides remote debugging functions such as reception and execution of the debugging commands. The monitor also provides a partial hardware emulation mechanism. It emulates only the hardware resources used by the remote debugging function (such as the interruption controller or interruption-handling table). The monitor provides the same interfaces as the real hardware, so it can work with any OSs running on PC/AT architectures.

The hardware resources used by the remote debugging functions are accessed by the original OS via the lightweight virtual machine monitor, so stability of the debugging environment can be assured because the real hardware remain in normal states even if the original OS works improperly because it has bugs. The other devices, however, especially high-throughput I/O devices (such as a SCSI controller or Ethernet controller) can be accessed directly by the original OS. This direct access enables I/O operations to be executed efficiently in this debugging environment. The direct access also makes it unnecessary for the monitor to emulate the high-throughput I/O devices, so this debugging environment can be used with various I/O devices without being customized.

This lightweight virtual machine monitor also provides a lightweight mechanism protecting memory regions. Even though x86 architecture provides two level memory protection mechanism, it can provide memory protection at three levels: that of the application on the original OS, that of the original OS, and that of the lightweight virtual machine monitor. This mechanism enables the remote debugging functions to continue working properly even when the application or the original OS executes illegal memory accesses to regions used by the monitor.

## 3. Performance Evaluation

We evaluated the performance of the lightweight virtual machine monitor by measuring the I/O performance of our original OS, the HiTactix, running on this monitor.

We ran the HiTactix OS on the following three PC/AT-compatible systems with 1.26GHz Pentium III[1] processors: real hardware, the lightweight virtual machine monitor, and a VMware[2] Workstation 4[2] for Linux. The VMware Workstation 4 is a typical example of a conventional virtual machine monitor for PC/AT compatible architecture that supports various I/O devices. On each of the systems running HiTactix, we executed a data-transfer application that reads 2MB data from three Ultra 160 SCSI disks at constant rates, splits them into 1024KB segments, and sends all segments via gigabit Ethernet using the UDP protocol. We measured the CPU load variations when we changed the data reading and transferring rates. Comparing the load variations measured for the different systems, we evaluated how much I/O performance was improved by using the lightweight virtual machine monitor instead of the VMware Workstation 4 and how much degradation of I/O performance are caused with real hardware when this monitor is used.

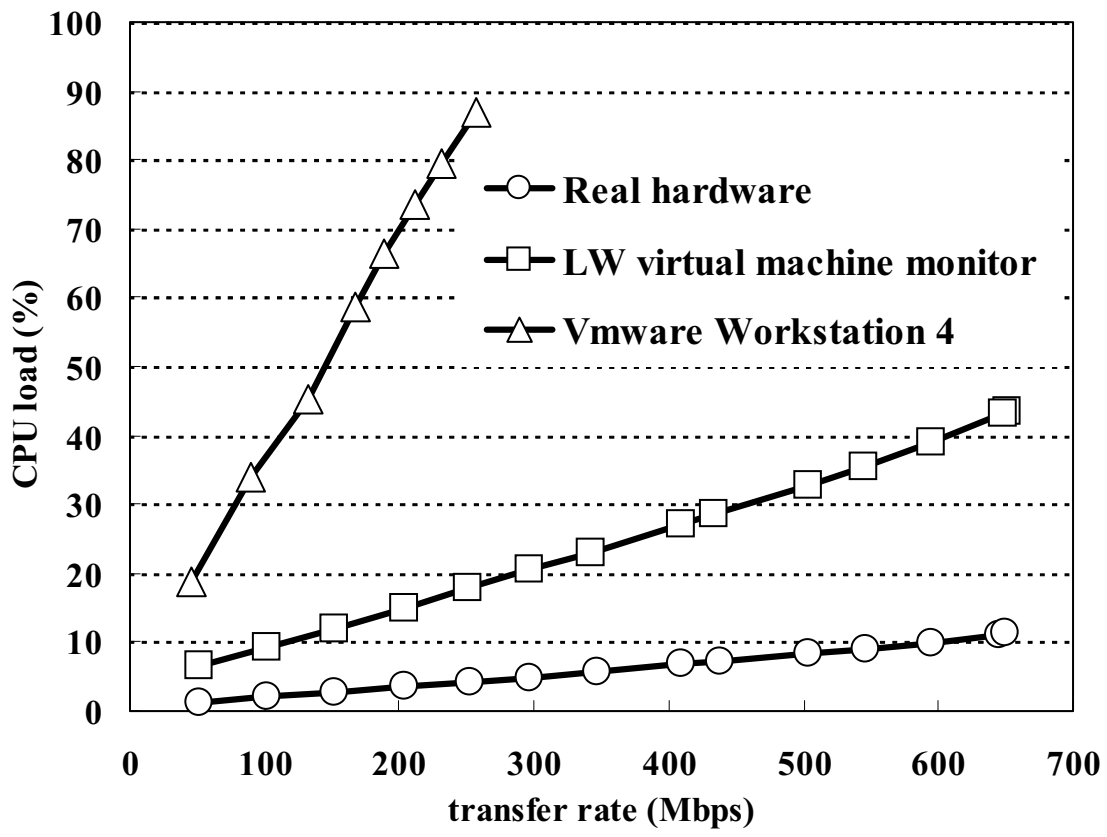

**Figure 3.1 Measured CPU load**

The measurement results plotted in Fig.3.1 show that our lightweight virtual machine monitor can transfer data 5.4 times as fast as a VMware Workstation 4 , but they also show that our monitor can transfer data at only about one fourth (26%) of the rate it can be transferred by real hardware.

## 4. Conclusion

This paper describes a novel OS debugging method and environment that can assure the stability of the debugging environment, be easily customized to a new OS or to a new I/O device, and assures efficient execution of I/O operations.

We evaluated the I/O performance provided by our proposed monitor and confirmed that the monitor can transfer data 5.4 times faster than a conventional virtual monitor.

---

1 Pentium III is a trademark of Intel Co. Ltd.

2 VMware is a trademark of VMWare Co. Ltd.